\documentclass[
twocolumn, 
showpacs,
prl, 
superscriptaddress] 
{revtex4}
\usepackage{graphicx}
\usepackage{color}
\usepackage{ulem}
\usepackage{verbatim}
\usepackage{amsmath}
\usepackage{amssymb}

\begin{document}

\title{Controlled rephasing of single collective spin excitations in a cold atomic quantum memory}%

\author{Boris Albrecht}
\affiliation{ICFO-Institut de Ciencies Fotoniques, Mediterranean Technology Park, 08860 Castelldefels (Barcelona), Spain}%

\author{Pau Farrera}
\affiliation{ICFO-Institut de Ciencies Fotoniques, Mediterranean Technology Park, 08860 Castelldefels (Barcelona), Spain}%

\author{Georg Heinze}
\affiliation{ICFO-Institut de Ciencies Fotoniques, Mediterranean Technology Park, 08860 Castelldefels (Barcelona), Spain}%

\author{Matteo Cristiani}
\affiliation{ICFO-Institut de Ciencies Fotoniques, Mediterranean Technology Park, 08860 Castelldefels (Barcelona), Spain}%

\author{Hugues de Riedmatten}
\affiliation{ICFO-Institut de Ciencies Fotoniques, Mediterranean Technology Park, 08860 Castelldefels (Barcelona), Spain}
\affiliation{ICREA-Instituci\'{o} Catalana de Recerca i Estudis Avan\c cats, 08015 Barcelona, Spain}%

\date{\today}

\begin{abstract}
We demonstrate active control of inhomogeneous  dephasing and rephasing  for single
collective atomic spin excitations (spin-waves) created by
spontaneous Raman scattering in a quantum memory based on cold
$^{87}$Rb atoms. The control is provided by a reversible external magnetic
field gradient inducing an inhomogeneous broadening of the atomic
hyperfine levels.  We  demonstrate experimentally that active rephasing preserves the
single photon nature of the retrieved photons. Finally, we show 
that the control of the inhomogeneous dephasing enables the
creation of time-separated spin-waves in a single ensemble, followed
by a selective readout in time. This is an important step towards
the implementation of a functional temporally multiplexed quantum
repeater node.
\end{abstract} 

\maketitle

Photonic quantum memories (QMs) \cite{Lvovsky2009, Hammerer2010,Bussieres2013} are
devices which can faithfully store and retrieve quantum
information encoded in photons. They are essential building blocks
for scalable quantum technologies involving photons, such as
linear optics quantum computing and long distance quantum
communication using quantum repeaters
\cite{Briegel1998,Duan2001,Sangouard2011}. QMs have been
implemented with single atoms and ions \cite{Ritter2012,
Moehring2007, Hofmann2012, Stute2012, Schug2013}, atomic vapors
\cite{Chou2005, Chaneliere2005, Eisaman2005, Choi2008,
Radnaev2010, Hosseini2011, Sprague2014, Bao2012, Bimbard2014, Nicolas2014}, and solid state
systems \cite{Riedmatten2008, Hedges2010, Clausen2011,
Saglamyurek2011,Rielander2014,Bernien2013}.

Atomic ensembles provide an efficient way of reaching the strong
interaction between matter and light required for the
implementation of quantum memories, without the need for high
finesse cavities. In addition, they give the possibility to
multiplex quantum information, which is desirable in several
applications. In particular, in quantum repeater architectures,
this alleviates the limitation due to the communication time
between the nodes \cite{Simon2007}. In atomic ensembles, quantum
information is stored as collective atomic spin excitations,
called spin-waves. Single spin-waves offer the important advantage
that they can be efficiently transferred to single photons in a
well defined spatio-temporal mode thanks to constructive
interference between the involved atoms. In 2001, Duan, Cirac,
Lukin and Zoller (DLCZ) proposed a protocol to implement a quantum
repeater using the heralded creation of spin-waves in an atomic
ensemble \cite{Duan2001}. Several demonstrations of the building
block of the protocol have been reported
\cite{Kuzmich2003,Chou2005,Radnaev2010} including functional
elementary segments of a quantum repeater \cite{Chou2007,Yuan2008}. Most of
these demonstrations however, used only a single spin-wave per
ensemble, although several ensembles have been already implemented
in the same atom trap \cite{Lan2009,Choi2010}.

A recent theoretical proposal has shown that the ability to
precisely control the quantum state of single spin-waves would open
new avenues for the realization of more efficient, temporally
multiplexed quantum repeater architectures \cite{Simon2010}. In
particular, it has been proposed that the implementation of
controlled dephasings and rephasings of single spin-waves would
allow the creation of several time-separated single spin-waves in
a single ensemble which could be selectively read out at different
times. Rephasing protocols have also been proposed and implemented
for optical collective atomic excitations, leading to the storage
and retrieval of 64 weak optical modes in a rare-earth doped crystal
\cite{Usmani2010}, with pre-determined storage times. The capability to control the dephasing of
spin-waves has also been used recently to implement a coherent optical pulse sequencer \cite{Hosseini2009}, efficient
light storage for bright \cite{Hosseini2011a} and weak coherent
pulses \cite{Hosseini2011} using the gradient echo memory
protocol \cite{Hetet2008a}, as well as storage of several temporal modes of bright pulses \cite{Hosseini2011a}.

In this paper we demonstrate active control of the inhomogeneous
dephasing and rephasing of single spin-waves stored in an atomic ensemble. A
cold $\mathrm{{}^{87}Rb}$ atomic ensemble QM is used to create
heralded single spin excitations using the DLCZ scheme. We
implement a controlled and reversible  spin inhomogeneous
broadening using a  magnetic field gradient \cite{Hetet2008a}, leading to
 spin-wave dephasing and rephasing at a controlled time. We infer the dephasing and rephasing of the spin-waves by
converting them into  single photons. We show that with this
ability, several temporally separated spin-waves can be stored and
read out selectively. This is the first enabling step towards the
implementation of a temporally multiplexed DLCZ quantum repeater,
following the proposal of \cite{Simon2010}.

\begin{figure}[h!]
\includegraphics[width=.48\textwidth]{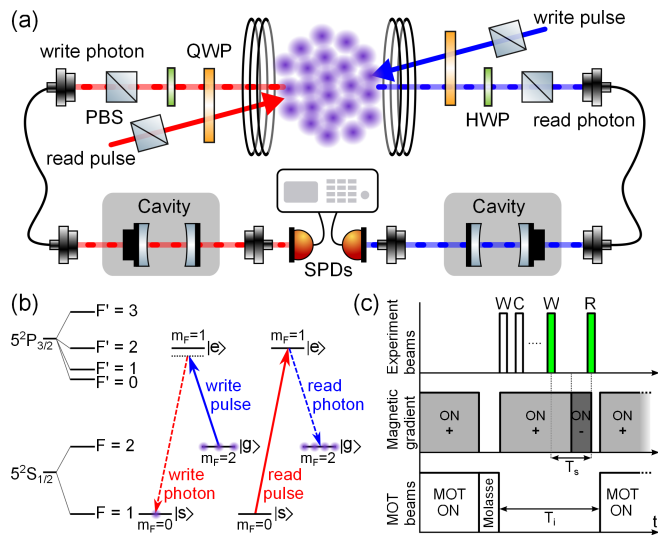}
\caption{(color online) (\textbf{a}) Schematic of the experimental setup. HWP: half-wave plate, QWP: quarter-wave plate, PBS: polarizing beam splitter, SPD: single photon detector. (\textbf{b}) Energy levels structure. (\textbf{c}) Experimental sequence timeline for the active rephasing experiment. W: write pulse, C: cleaning stage, R: read pulse, $T_s$: storage time, $T_i$: interrogation time}
\label{Figure1}
\end{figure}

To implement a DLCZ QM, we start by an optically thick ensemble of
$N$ identical atoms exhibiting a $\Lambda$-type level scheme with two
metastable ground states $\left|g\right\rangle$ and
$\left|s\right\rangle$ and an excited state $\left|e\right\rangle$
(see Fig.~\ref{Figure1}(b)). All the atoms are initially prepared
in $\left|g\right\rangle$. A weak, off-resonant write pulse on the
$\left|g\right\rangle \rightarrow \left|e\right\rangle$ transition
probabilistically creates a spin-wave heralded by a Raman
scattered write photon. The write photons and the associated
spin-waves are described by a two-mode squeezed state as
\begin{equation}
\left|\phi\right\rangle = \sqrt{1-p}(\left|0_w\right\rangle \left|0_s\right\rangle + \sqrt{p}\left|1_w\right\rangle \left|1_s\right\rangle + p\left|2_w\right\rangle \left|2_s\right\rangle + o(p^{3/2})),
\label{eq1}
\end{equation}
where $p$ is the probability to create a spin-wave per trial in
the detection mode, and $w$ and $s$ stand for write photon and
spin-wave respectively. Upon detection of a write photon, the
state of the associated spin-wave to first order is given by
\begin{equation}
\left|1_s\right\rangle=\left| \psi(0) \right\rangle =
\frac{1}{\sqrt{N}} \sum_{j=1}^{N}
e^{i\textbf{x}_j\cdot(\textbf{k}_W - \textbf{k}_w )} \left|g_1
\ldots s_j \ldots g_N\right\rangle, \label{eq2}
\end{equation}
where $\textbf{x}_j$ is the position of atom $j$, and $\textbf{k}_W$ and $\textbf{k}_w$
 are the wavevectors of the write pulse and write photonic mode respectively.
The spin-wave can be read out at a later time with a
counterpropagating read pulse resonant with the
$\left|s\right\rangle \rightarrow \left|e\right\rangle$
transition. This converts the atomic excitation into a single read
photon. Thanks to  collective interference of all contributing atoms, the
read photon is emitted in a well defined spatial mode given by the
phase matching condition $\textbf{k}_r = \textbf{k}_R +
\textbf{k}_W - \textbf{k}_w$, where $\textbf{k}_R$ and
$\textbf{k}_r$ are the wavevectors of the read pulse and read
photonic mode respectively. The retrieval efficiency is defined
as $\eta_{ret}=p_{w,r}/p_w$, where $p_{w,r}$ is the
probability per trial to detect a coincidence between write and read
photons, and $p_w$ is the probability to detect a write photon.

For the active rephasing experiment, we apply a magnetic field
gradient to the ensemble during the storage of the spin-waves.
This creates an inhomogeneous broadening
of the hyperfine levels, due to the spatially dependent Zeeman
shift and leads to a rapid dephasing of the spin-wave, with no
further spontaneous rephasing. In this case, the state of the
spin-wave evolves as
\begin{multline}
\left| \psi(t) \right\rangle = \frac{1}{\sqrt{N}} \sum_{j=1}^{N} e^{i\int_0^t \Delta \omega_j(t') dt' + i(\textbf{x}_j+\textbf{v}_j t)\cdot (\textbf{k}_W - \textbf{k}_w)}\\
\left|g_1 \ldots s_j \ldots g_N\right\rangle, \label{eq3}
\end{multline}
where $\Delta \omega_j(t)$ is the relative
detuning of the state associated with atom $j$ in
$\left|s\right\rangle$,  which is proportional to the magnetic gradient
amplitude, and $\textbf{v}_j$ is the velocity of atom $j$.

 If we
reverse the magnetic gradient at a time $T_{rev}$, a rephasing will
be induced when $\int_0^{T_{rev}} \Delta \omega_j(t') dt'
+ \int_{T_{rev}}^t \Delta \omega_j(t') dt'=0$, which happens at a
time $2T_{rev}$ for a symmetric reversal, e.g. if $\Delta
\omega_j(t\leq T_{rev}) = -\Delta \omega_j(t\geq T_{rev})$.
The retrieval efficiency at a time $t$ is proportional to the
overlap of the state at that time with the initial state:
$\eta_{ret}(t) \propto
\left|\left\langle\psi(0)|\psi(t)\right\rangle\right|^2
$ \cite{Zhao2009}. Therefore, we use
$\eta_{ret}$ to monitor the spin-waves dephasing and rephasing in the following.

The experimental setup is shown in Fig.~\ref{Figure1}(a) \cite{Albrecht2014}. We use an
ensemble of laser cooled $\mathrm{^{87}Rb}$ atoms loaded in a
magneto optical trap (MOT), with a longitudinal magnetic gradient
of $20\mathrm{G}/\mathrm{cm}$ along the probing axis. We address
the $D_2$ line, which is resonant with light at $780\,\mathrm{nm}$.
The cooling laser, red detuned from the
$\left|F=2\right\rangle \rightarrow \left|F'=3\right\rangle$
transition, and the repumper laser, resonant with the
$\left|F=1\right\rangle \rightarrow \left|F'=2\right\rangle$
transition, will be referred to as the trapping beams.  All the atoms are initially optically
pumped in the $\left|g,m_F=2\right\rangle$ Zeeman sublevel,
permitting us to create a magnetically sensitive spin-wave, which
is necessary for this experiment. The write pulse is red detuned
by $40\,\mathrm{MHz}$ from the $\left|g\right\rangle \rightarrow
\left|e\right\rangle$ transition and has a duration of
$16\,\mathrm{ns}$. With these parameters, a peak power of 590
$\,\mu W$ leads to a write photon detection probability $p_w$ of
$1\%$. The read pulse, counterpropagating with the write pulse, is
resonant with the $\left|s\right\rangle \rightarrow
\left|e\right\rangle$ transition and has a duration of
$20\,\mathrm{ns}$. A peak power of $570\,\mu W$ maximizes the
retrieval efficiency. 
The polarization of the write and read pulses in the frame of the atoms is $\sigma^-$ and
$\sigma^+$ respectively, while the detected write and read photons are $\sigma^+$ and $\sigma^-$ polarized. 
We set an angle of $0.95^{\circ}$ between the write/read
pulses axis and the photons detection axis, to spatially
separate the classical pulses from the write and read photons.
The write and read photons are spectrally filtered by
identical monolithic Fabry-Perot cavities with $20\,\%$ total
transmission (including cavity transmission, 
subsequent fiber coupling and spectrum mismatch between the single photons and the cavity mode), before detection by single photon
detectors (SPDs) with $43\,\%$ efficiency.

The experimental sequence for the active rephasing experiment is
shown in Fig.~\ref{Figure1}(c). After a MOT loading
phase of $15\,\mathrm{ms}$, we switch off the MOT gradient while the trapping
beams are kept on during a molasses ($1.6\,\mathrm{ms}$) phase.
Afterwards, we perform optical pumping for $10\,\mathrm{\mu s}$. The measured optical depth at this stage is $6 \pm 1$. The magnetic
gradient is then switched on again before the beginning of an
interrogation period of up to $660\,\mathrm{\mu s}$
during which a train of up to $200$ write pulses is sent. For each
trial, if no write photon is detected, a cleaning stage sets the
memory back into its initial state. It consists of an optical pumping
pulse, which is $\sigma^+$ polarized and
resonant with the $\left|F=2\right\rangle \rightarrow
\left|F'=3\right\rangle$ transition, and read light. If instead a
write photon is detected, we reverse the magnetic gradient and send a read pulse after a programmable delay.
This leads to the end of the interrogation of the current ensemble
and to the loading of a new one with a repetition rate of $59\,\mathrm{Hz}$.

\begin{figure}
\includegraphics[width=.48\textwidth]{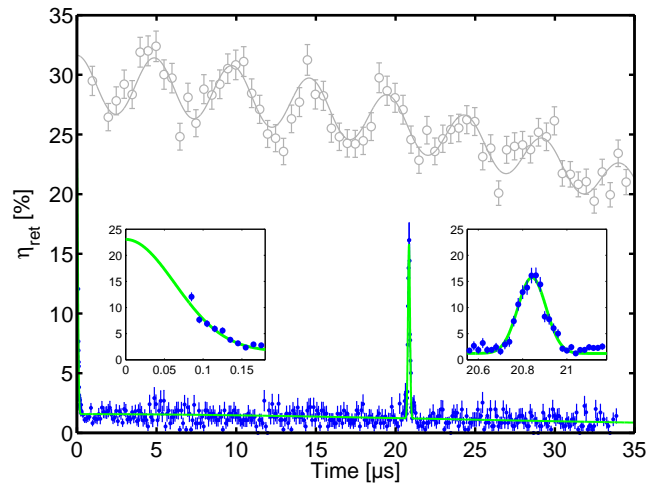}
\caption{(color online) Retrieval efficiency vs. readout time for $p_w = 1\%$. (grey open circles) Standard DLCZ experiment. The oscillations are due to imperfect optical pumping. (grey line) Fit of the experimental data. (blue dots) Active rephasing experiment. (green line) Fit of the experimental data.}
\label{Figure2}
\end{figure}

We now present the experimental results. We start by measuring the
retrieval efficiency $\eta_{ret}$ as a function of storage
time, for a standard DLCZ experiment, i.e. when no magnetic gradient
is applied during the interrogation period. The results are shown
in Fig.~\ref{Figure2} (open circles). We observe an
oscillation which can be attributed to a beating between two different classes of spin-waves due to imperfect optical pumping into the 
$\left|g,m_F=2\right\rangle$ state \cite{Zhao2009a,SupMat}.
The overall decay in retrieval efficiency has a time constant of $57\pm 1\,\mathrm{\mu s}$, limited by atomic motion \cite{Zhao2009, Nicolas2014} and spurious magnetic field gradients \cite{Felinto2005}. 
We then switch on the gradient during the interrogation period. In
that case, we observe a rapid dephasing of the spin-wave at short storage times (see
left inset), followed by background noise for a period of about
$20\,\mu\mathrm{s}$. If the gradient is reversed after the
detection of a write photon, we observe a rephasing,
witnessed by a pronounced increase in retrieval efficiency. The measured
efficiencies after the rephasing are again at the noise level. The
reversal instruction is sent $3\,\mu\mathrm{s}$  after the write pulse, but due to the
temporal response of the coils driving circuit, the rephasing
occurs at $20.84\,\mu\mathrm{s}$ (see right inset). The full
width at half maximum of the rephasing peak is
$150\pm 3\,\mathrm{ns}$, and its position can be precisely modified by changing
the magnetic gradient reversal time \cite{SupMat}.

   The retrieval efficiency at the rephasing peak
 is about $60\%$ of the retrieval efficiency in the
standard DLCZ experiment for the same storage time \cite{SupMat}.
Several effects may explain this reduction.  First, slow fluctuations of the current in the coils generating the magnetic field gradient can change the position of the rephasing peak during the measurement time, which will decrease the efficiency  for a given read-out time. Moreover, time varying magnetic field gradients over the interrogation time lead to fluctuations of the rephasing time depending on the write photons detection time. Finally, the quality of the optical pumping may be degraded during the interrogation period because of the presence of the magnetic field gradient. This  decreases the optical depth and therefore the efficiency.  The
experimental data are fitted based on equation \eqref{eq3},
corrected for the memory lifetime and the temperature drift of the
read photons filtering cavity during the measurement time (see
\citep{SupMat}). The signal to noise ratio (SNR), calculated as
the maximum value of the fit at the rephasing time divided by the
mean of the background, is $13.3 \pm 0.9$. 

\begin{figure}
\includegraphics[width=.48\textwidth]{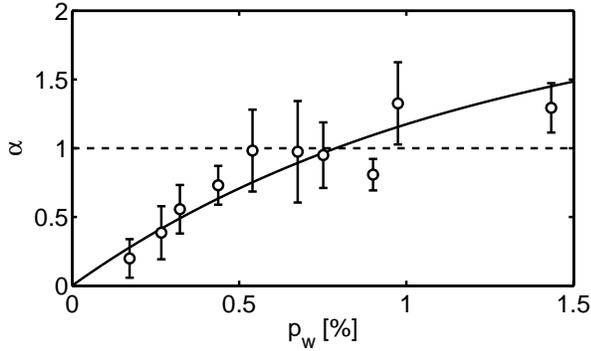}
\caption{Antibunching parameter as a function of write photon detection probability. Open circles: experimental data. The dashed line represents the limit for classical states ($\alpha \geq 1$).}
\label{Figure3}
\end{figure}

Next, we investigate the single photon nature of the rephased read photons.  For this measurement, we modified the setup by
sending the read photons through a balanced fiber beamsplitter
connected to two SPDs. We measured the antibunching parameter
$\alpha$ of the read photons \cite{Grangier1986} defined as
\begin{equation}
\alpha = \frac{p_{w,r_1,r_2}\cdot p_w}{p_{w,r_1}\cdot p_{w,r_2}},
\label{eq4}
\end{equation}
where $p_{w,r_1,r_2}$ is the probability to measure a triple
coincidence between a write photon and both read photons
detections, and $p_{w,r_1}, p_{w,r_2}$ are the probabilities to
measure a coincidence between a write photon and either one of the
read photon detections. The measured values of $\alpha$ as a
function of $p_w$ in the case of the active rephasing experiment
are shown in Fig.~\ref{Figure3}. An antibunching parameter
lower than 1 is a proof of  non-classicality, 1 corresponding to coherent
states and 0 to perfect single photons. We expect to be in the
single photon regime for low excitation probabilities.  We observe
values below 1 within the error for $p_w$ up to $0.5\,\%$, with
values as low as $0.20 \pm 0.14$ for $p_w = 0.17\%$. The data are fitted
using the formula $\alpha = 2p(2c(1+p)-p)/(c(1+p))^2 $, where $c$ is the proportionality factor between the real and ideal $g^{(2)}_{w,r}$, the cross-correlation function between write and read photons (see \cite{SupMat}). 

\begin{figure}
\includegraphics[width=.48\textwidth]{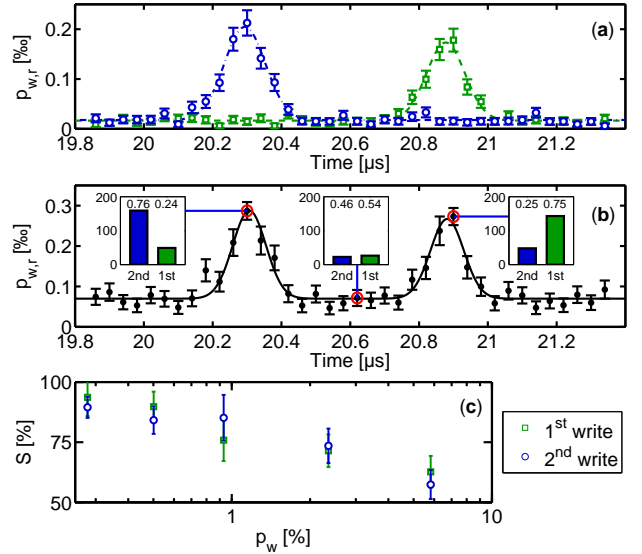}
\caption{(Color online) Temporally separated spin-waves.  (\textbf{a}) Single rephasing case for $p_w = 1\%$: first write pulse in green, second write pulse in blue. (\textbf{b}) Both pulses sent in black.  The histograms  display the relative weight of each peak, at the circled points. (\textbf{c}) Selectivity as a function of $p_w$ for each rephasing.}
\label{Figure4}
\end{figure}

It has been predicted that the controlled dephasing and rephasing
of single collective excitations as demonstrated in this paper should enable the creation
of multiple time-separated spin-waves that can be selectively
read-out in time \cite{Simon2010}. To confirm this prediction, we send two write
pulses separated by $600 \,\mathrm{ns}$, first independently and
then conjointly before performing the readout around the rephasing
time. Fig.~\ref{Figure4} (a) shows the coincidence probabilities
per trial $p_{w,r}$ when each write pulse is send independently.
 The readout time is defined as the time between the
first write pulse and the read pulse. As expected, the spin-wave
created by the first write pulse rephases later and vice versa.
Fig.~\ref{Figure4} (b) shows the coincidence probability when both
write pulses are sent conjointly. The values at the maximum of the
rephasing peaks are similar, when background-subtracted. However,
the background  probability outside the rephasing peaks is higher in this case. This can be
explained by expressing the coincidence probability in the
background outside the rephasings as $p_{B} = p_w\cdot p_r$.
Noting that $p_r \propto p_w$ \cite{Chen2006}, one gets $p_{B} \propto
p_w^2$. In the case where we send two write pulses, the write
detection probability $p_{2w}$ is twice the one of the single
write pulse case $p_w$. Therefore, the background probability for
two write pulses becomes $p_{B}^{(2)} \propto p_{2w}^2 = 4*p_w^2$. This is compatible with the measured value of $4.1 \pm 0.3$. 

To investigate the cross-talk between the two spin-waves, we construct the histograms shown in the insets by performing start-stop
measurements. The starts are write photon detections, and the
stops are read photon detections. The two contributions are equally
weighted in the noise region, but on each rephasing peak, we
detect a significant imbalance. This shows that mostly only one spin-wave
rephases at a time. To quantify the process, we calculate the relative weight of each
peak, which we call selectivity: $S(i) = p_{C,i}/\sum_k p_{C,k}$,
with $p_{C,i}$ the probability to detect a coincidence in the
binning corresponding to the peak number $i$. Its value depends on
the rephasing SNR which varies with $p_w$ (see Fig.~\ref{Figure4}
(c) and \cite{SupMat}). For the results shown in Fig.~\ref{Figure4}
(b) we find $S$ = $(S(1)+S(2))/2 = 76 \pm 6 \%$. For lower
$p_w = 0.28\%$, we obtain a selectivity of $92 \pm 4 \%$. 

The results of Fig.~\ref{Figure4} show that when several
spin-waves are created, the rephasing efficiency
 of each one remains the same, but the background noise increases.
 With the current status of the experiment, this limits the benefit of multiple temporal mode
 storage, since the excitation probability needs to be reduced for
 each pulse in order to keep the same SNR. 
However, this issue is addressed in the scheme proposed in
\cite{Simon2010}. A possible solution would be to build a low
finesse cavity resonant with the write photons but invisible to
the read photons around the QM. This would decrease the proportion
of unwanted spin-waves created per write pulse by increasing the
proportion of write photons emitted in the cavity mode. Therefore, the noise would decrease by a factor corresponding to the
cavity finesse.

In conclusion, we demonstrated active rephasing of a single
spin-wave at a controllable time by inverting the polarity of an
external inhomogeneous broadening created by a magnetic gradient.
We showed that in this case, the retrieved photons still exhibit
antibunching, which proves that active rephasing preserves single
photons statistics. Finally, we demonstrated experimentally that this
technique enables the creation of multiple time-separated spin-waves
that can be read-out with high selectivity in time. These results pave
the way towards the realization of a temporally multiplexed
DLCZ-type quantum repeater node.

\section{Acknowledgements}
We acknowledge
financial support by the ERC starting grant QuLIMA and by the Spanish
Ministry of Economy and Competitiveness (MINECO) and the Fondo
Europeo de Desarrollo Regional (FEDER) through grant
FIS2012-37569. P.F. acknowledges the International PhD-fellowship program "la Caixa"-Severo Ochoa @ICFO.

\textit{ Note added}.--  Related works demonstrating spin echoes  at the single excitation level in  rare-earth doped crystals \cite{Jobez2015} and cold atomic ensembles \cite{Rui2015} have recently been reported. 

\bibliographystyle{prsty}

\end{document}